\documentclass[a4paper]{article}
\usepackage{INTERSPEECH2022}
\usepackage{amsmath,graphicx}
\usepackage[cmintegrals]{newtxmath}
\usepackage{bm}
\usepackage{algorithmicx,algorithm}
\usepackage{booktabs}
\usepackage{multirow}
\usepackage{graphicx}
\usepackage{stfloats}
\usepackage{subfig}
\usepackage{float}
\usepackage{hyperref}
\hypersetup{hidelinks}

\def\equationautorefname#1#2\null{
	Eq. (#2\null)
}

\usepackage{verbatim}

\usepackage{hyperref}
\hypersetup{hidelinks}

\def\equationautorefname#1#2\null{
	Eq. (#2\null)
}

\title{Wav2vec-S: Semi-Supervised Pre-Training for Low-Resource ASR}
\name{Han Zhu$^{1,2}$ \thanks{This work is partially supported by the National Key Research and Development Program of China (No. 2020AAA0108002).}, Li Wang$^{1}$, Jindong Wang$^{3}$, Gaofeng Cheng$^{1}$, Pengyuan Zhang$^{1,2}$, Yonghong Yan$^{1,2}$}
\address{$^1$Key Laboratory of Speech Acoustics and Content Understanding, Institute of Acoustics CAS, China \\
$^2$ University of Chinese Academy of Sciences, China   \qquad  $^3$ Microsoft Research Asia, China}
\email{zhuhan@hccl.ioa.ac.cn}

\begin{document}

\maketitle
\begin{abstract}
Self-supervised pre-training could effectively improve the performance of low-resource automatic speech recognition (ASR). However, existing self-supervised pre-training are \emph{task-agnostic}, i.e., could be applied to various downstream tasks. Although it enlarges the scope of its application, the capacity of the pre-trained model is not fully utilized for the ASR task, and the learned representations may not be optimal for ASR. In this work, in order to build a better pre-trained model for low-resource ASR, we propose a pre-training approach called \emph{wav2vec-S}, where we use \emph{task-specific} semi-supervised pre-training to refine the self-supervised pre-trained model for the ASR task thus more effectively utilize the capacity of the pre-trained model to generate task-specific representations for ASR. Experiments show that compared to wav2vec 2.0, wav2vec-S only requires a marginal increment of pre-training time but could significantly improve ASR performance on in-domain, cross-domain and cross-lingual datasets. Average relative WER reductions are 24.5\% and 6.6\% for 1h and 10h fine-tuning, respectively. Furthermore, we show that semi-supervised pre-training could close the representation gap between the self-supervised pre-trained model and the corresponding fine-tuned model through canonical correlation analysis.
\end{abstract}

\section{Introduction}
\label{sec:intro}

The performance of automatic speech recognition (ASR) heavily relies on the amount of labeled data, which is costly and not available in many low-resource scenarios.
To alleviate this issue, the self-supervised learning~\cite{baevski2020wav2vec,liu2021tera,yang2021superb,deng21b_interspeech,baevski2022data2vec,gao2021data,misra2021comparison} can be used to build the self-supervised pre-trained model with massive unlabeled data. However, since the self-supervised pre-training is \emph{task-agnostic}, the capacity of the pre-trained model is not fully utilized for ASR. And the representations of the self-supervised pre-trained model may not be optimal for ASR \cite{pasad2021layer}. As an alternative, conventional transfer learning approaches \cite{bell2020adaptation,li2017large,zhu2019accent,meng2020vector,zhu2020domain,hou2021exploiting} typically build a supervised pre-trained model with labeled data in the high-resource domain. Since the labeled data provide task-specific information, the supervised pre-trained model is \emph{task-specific} to ASR. However, since a considerable amount of unlabeled data are unused, the supervised pre-trained model could be unsatisfactory in performance.

In order to build a better pre-trained model for ASR, we propose a simple pre-training pipeline: \emph{wav2vec-S}, which uses both labeled and unlabeled data to learn the ASR task-specific representations.
Specifically, on the basis of the \emph{task-agnostic} \emph{self-supervised} pre-training, we further conduct \emph{task-specific} \emph{semi-supervised} pre-training to learn task-specific representation. The reason we use semi-supervised pre-training instead of supervised pre-training is that the amount of labeled data is limited. Since the unlabeled data in semi-supervised pre-training is utilized through pseudo-labeling \cite{chen2020semi,likhomanenko2020slimipl,higuchi2021momentum,park2020improved}, the semi-supervised pre-training is also learning task-specific representations. The two steps in wav2vec-S, i.e., self-supervised and semi-supervised pre-training, are loosely coupled. Thus the same strategy for semi-supervised pre-training can be used on the basis of different self-supervised pre-training approaches.

Experiments show that wav2vec-S consistently improves the self-supervised model thus could act as the alternative to the vanilla self-supervised model for the downstream ASR task.
Moreover, we performed detailed ablation studies for the semi-supervised pre-training step in wav2vec-S and the main conclusions are as follows:
\begin{itemize}
    \setlength{\itemsep}{0ex} 
    \item Semi-supervised pre-training can improve the performance and generalization of the self-supervised pre-trained model, i.e., improvements on in-domain, cross-domain and cross-lingual datasets.
    \item Character-level supervision is better than phone-level for monolingual semi-supervised pre-training even on a cross-lingual downstream dataset, which could alleviate the efforts to generate the phoneme transcriptions.
    \item Monolingual semi-supervised pre-training has a trade-off between performance of the source language and other languages. With more training updates, the model would become more language-specific, and the cross-lingual generalization ability is thus degraded.
    \item The semi-supervised pre-training step costs much less time than self-supervised pre-training. Thus wav2vec-S only has a marginal increment of pre-training time than vanilla self-supervised pre-training.
    \item Semi-supervised pre-training effectively improves different self-supervised pre-trained models, e.g., wav2vec 2.0 \cite{baevski2020wav2vec}, data2vec \cite{baevski2022data2vec}.
    \item  We analyze the representation similarity before and after fine-tuning for pre-trained models with canonical correlation analysis (CCA), and show semi-supervised pre-training closes the representation gap between the pre-trained and fine-tuned models.
\end{itemize}

\section{Related works}
\label{sec:related}

The idea to adapt the task-agnostic self-supervised pre-trained model to an ASR task-specific pre-trained model is also explored in other works. \cite{wang2021self} forces the
model to concentrate on ASR-related information by adding the self-supervised losses on intermediate layers. Other works \cite{wang2021unispeech,bai2021joint,zhang2021xlst,chen2021speech} utilize labeled data to inject ASR task information into the pre-trained model. Our work belongs to this category. Among them, Unispeech \cite{wang2021unispeech} uses multi-task learning to conduct semi-supervised pre-training, where contrastive loss is used on the unlabeled data and CTC loss is used on the labeled data. JUST \cite{bai2021joint} jointly optimizes two self-supervised losses and a supervised RNN-T loss. XLST \cite{zhang2021xlst} uses supervised training as the initialization and then conducts self-training on the unlabeled data. In our work, CTC loss is used on both labeled and unlabeled data, where the ground-truth labels are used for labeled data and pseudo labels are used for unlabeled data. Since previous work mostly conducts task-specific pre-training from scratch, substantial training time is required for each task. In this work, we treat semi-supervised pre-training as the task-specific refinement of the self-supervised pre-training. Thus, it can benefit from the initialization of the self-supervised pre-trained model for faster convergence. Concurrent works \cite{hwang2021large,zhang2021bigssl} also explored the combination of self-supervised pre-training and semi-supervised learning, where \cite{hwang2021large} focused on the domain adaptation and \cite{zhang2021bigssl} focused on the large-scale applications.

\section{Proposed approach}
\label{sec:proposed}

\subsection{Problem Formulation}

We denote the pre-training and fine-tuning dataset as the source domain $S$ and target domain $T$. 
Suppose the source domain consists of an unlabeled dataset $\mathbb{U}^{S}=\left\{\mathbf{x}_{1}^{S}, \ldots, \mathbf{x}_{N}^{S}\right\}$ and a labeled dataset $\mathbb{L}^{S}=\left\{\left(\mathbf{x}_{1}^{S}, \mathbf{y}_{1}^{S}\right), \ldots,\left(\mathbf{x}_{M}^{S}, \mathbf{y}_{M}^{S}\right)\right\}$, where $M \leq N$.
These two datasets are used for pre-training, where the self-supervised pre-training uses $\mathbb{U}^{S}$ and semi-supervised pre-training uses both $\mathbb{U}^{S}$ and $\mathbb{L}^{S}$. Fine-tuning is performed on an additional target domain labeled dataset $\mathbb{L}^{T}=\left\{\left(\mathbf{x}_{1}^{T}, \mathbf{y}_{1}^{T}\right), \ldots,\left(\mathbf{x}_{O}^{T}, \mathbf{y}_{O}^{T}\right)\right\}$

\subsection{Model Structure}

We adopt the model structure in wav2vec 2.0 \cite{baevski2020wav2vec}, which is shown in the first step of \autoref{fig:wav2vec-S}. A convolutional feature encoder is used to map the input raw audio $\mathcal{X}$ to higher-level latent speech representations $\mathcal{Z}$, which is then fed to the transformer context network to build context representations $\mathcal{C}$. During pre-training or fine-tuning, the mask module masked a proportion of the feature encoder outputs $\mathcal{Z}$ to $\mathcal{Z}^{'}$ before feeding them into the context network. Note that the masked dimension is only the time dimension during self-supervised pre-training, while it consists of both time and channel dimensions during semi-supervised pre-training and fine-tuning like SpecAugment \cite{park2019specaugment}. The masked time steps are denoted as gray color in \autoref{fig:wav2vec-S}. 
\begin{comment}
During self-supervised pre-training, an additional quantization module will discretize the masked time steps of feature encoder outputs $\mathcal{Z}$ to a finite set of quantized representations $\mathcal{Q}$, which are used as targets in the self-supervised objective.

During pre-training, the main objective is a contrastive loss $\mathcal{L}_{c}$, which encourage the model to identify the true quantized latent speech representation $q^{+}$ in a set of quantized candidate representations $\{q^{+},q^{-}, \ldots, q^{-}\}$. Furthermore, the wav2vec 2.0 objective is augmented by a diversity loss $\mathcal{L}_{d}$, which encourages the equal use of each quantized representation. We omit the details of $\mathcal{L}_{d}$ for simplicity. Overall, the wav2vec 2.0 objective is 
\begin{equation}
\label{equ:wav2vec}
\mathcal{L}_{w}=\mathcal{L}_{c}+ \alpha \mathcal{L}_{d},
\end{equation}
where $\alpha$ is a tuned hyperparameter.

During fine-tuning, the quantization module is discarded and the mask module is optional for data augmentation. A random initialized linear layer is added after context representations to project the dimension to the vocabulary size. The resulting model is then optimized using task-specific loss.
\end{comment}

\subsection{Wav2vec-S}
\label{sec:wav2vec-S}

\begin{figure}[!t]
	\centering
	\includegraphics[width=1.0\columnwidth]{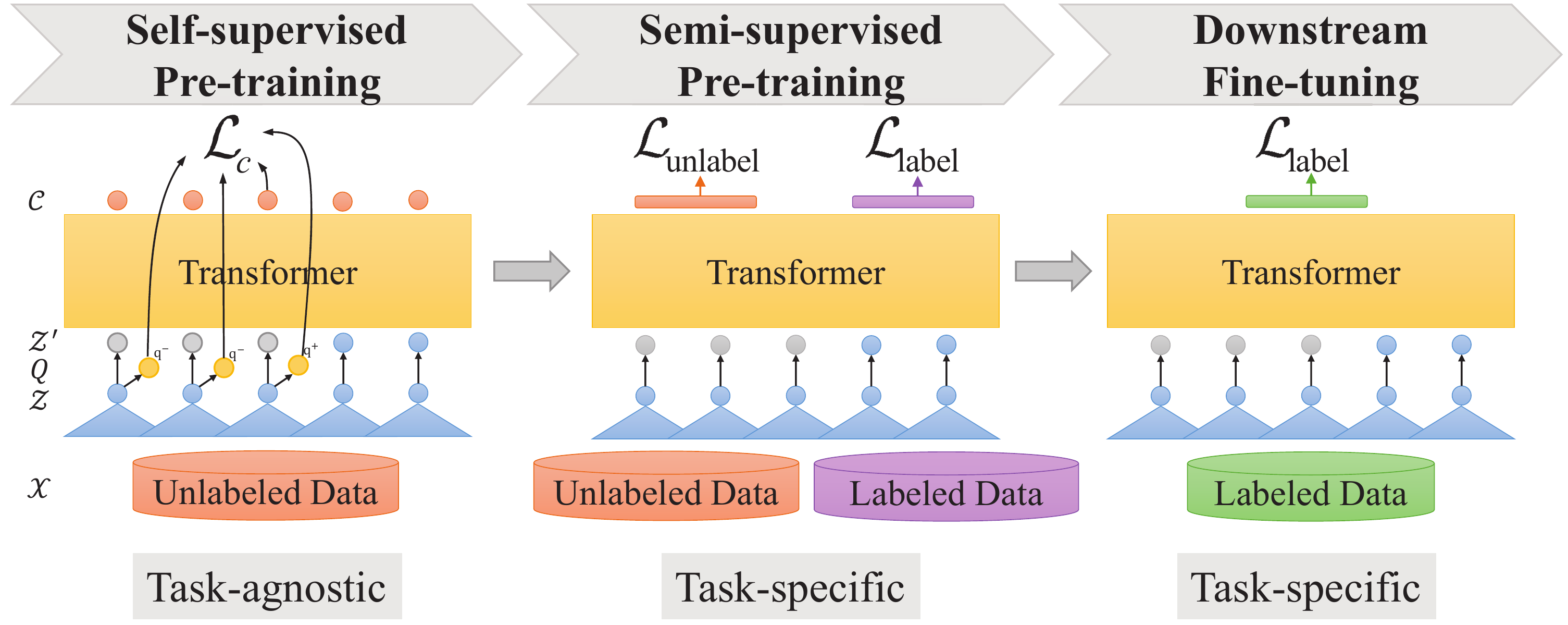}
	\caption{Illustration of the wav2vec-S procedure.} 
	\label{fig:wav2vec-S}
\end{figure}

We illustrate the wav2vec-S procedure in \autoref{fig:wav2vec-S}. The pre-training consists of two steps.
Firstly, self-supervised pre-training is performed on the unlabeled source dataset $\mathbb{U}^{S}$ using the self-supervised loss. Then, semi-supervised pre-training is applied on both labeled source dataset $\mathbb{L}^{S}$ and unlabeled source dataset $\mathbb{U}^{S}$.
The total loss for semi-supervised learning is:
\begin{equation}
\label{equ:wav2vec-S}
\mathcal{L}_{\text{semi}}=\mathcal{L}_{\text{label}} + \lambda \mathcal{L}_{\text{unlabel}},
\end{equation}
where $\mathcal{L}_{\text{label}}$ and $\mathcal{L}_{\text{unlabel}}$ denotes the loss for labeled and unlabeled data, respectively. $\lambda$ is the hyperparameter to be tuned, which is fixed to 1 in this work.

Specifically, for ASR, during semi-supervised pre-training, both labeled and unlabeled losses are CTC \cite{graves2006connectionist}. For labeled data, it is straightforward to compute the CTC loss as:
\begin{equation}
\label{equ:label}
\mathcal{L}_{\text{label}} = -\mathbb{E}_{\mathbf{x}, \mathbf{y} \sim p(\mathbf{x}, \mathbf{y})} \log p_{\mathbf{\theta}}(\mathbf{y} \mid a(\mathbf{x})), (\mathbf{x}, \mathbf{y}) \in \mathbb{L}^{S}
\end{equation}
where $(\mathbf{x}, \mathbf{y})$ is the sample-label pair, $p(\mathbf{x}, \mathbf{y})$ is the distribution of samples from $\mathbb{L}^{S}$, $\mathbf{\theta}$ is the model parameter, and $a(\cdot)$ is the augmentation function.

However, for the unlabeled data, since the ground-truth labels are not available, pseudo labels are used instead:
\begin{equation}
\label{equ:unlabel}
\mathcal{L}_{\text{unlabel}} = -\mathbb{E}_{\mathbf{x} \sim p(\mathbf{x})} \log p_{\mathbf{\theta}}(\hat{\mathbf{y}} \mid a(\mathbf{x})), \mathbf{x} \in \mathbb{U}^{S}
\end{equation}
where $\hat{\mathbf{y}}$ denotes the pseudo label which is generated through:
\begin{equation}
\label{equ:pseudolabel}
\hat{\mathbf{y}}=\underset{\mathbf{y}}{\operatorname{argmax}} \log p_{\mathbf{\theta}}(\mathbf{y} \mid \mathbf{x}),
\end{equation}
where $\operatorname{argmax}$ denotes the greedy decoding, which first takes the maximum probability tokens in each frame and then removes repeated and blank tokens.
Note that the pseudo labels are generated using the up-to-date model $\theta$ on-the-fly as in \cite{chen2020semi}.

After pre-training, the task-specific labeled loss $\mathcal{L}_{\text{label}}$ is also used to fine-tune the pre-trained model on the target domain labeled dataset $\mathbb{L}^{T}$.

\section{Experiments}
\label{sec:experiments}

\subsection{Corpus}
The pre-training (source) dataset is LibriSpeech \cite{panayotov2015librispeech}, where the 100h clean subset is used as the labeled dataset $\mathbb{L}^{S}$ and the other 860h is the unlabeled dataset $\mathbb{U}^{S}$. As for fine-tuning (target) labeled datasets $\mathbb{L}^{T}$, Wall Street Journal (WSJ) and the US accented part of AESRC \cite{shi2021accented} is used as the in-domain dataset since they are all read datasets. To verify the generalization ability, conversational dataset SwitchBoard (SWBD) \cite{godfrey1993switchboard} and lecture dataset TED-LIUM3 (TED) \cite{hernandez2018ted} are used as cross-domain datasets. Moreover, Mandarin Chinese dataset AISHELL-1 \cite{bu2017aishell} and French dataset from Common Voice (CV French) \cite{ardila2020common} are used as cross-lingual datasets. We concentrate on the low-resource scenario, thus randomly selecting 1h or 10h subset of the above datasets for fine-tuning.

\subsection{Implementation Details}
All experiments are conducted with fairseq \cite{ott2019fairseq}. For self-supervised pre-training, we use the open-source wav2vec 2.0 or data2vec base model pre-trained on Librispeech 960h. For semi-supervised pre-training, we use gradient accumulation to achieve an effective batch size of 25.6m samples. The maximum learning rate is $3\times10^{-5}$, and the tri-state learning rate schedule from \cite{baevski2020wav2vec} is used.
And the convolutional feature encoder is fixed during training. For fine-tuning, batch size and learning rate are the same with semi-supervised pre-training. Apart from the convolutional feature encoder, the transformer context network is also fixed for the first 10k updates. The total training updates for 10h and 1h fine-tuning are 20k and 13k, respectively. Beam-search decoding with a dataset-specific 4-gram language model is used for evaluation. 

\begin{table*}[htbp]
  \centering
  \caption{1h and 10h fine-tuning with different pre-training approaches.}
    \resizebox{.9\textwidth}{!}{
    \begin{tabular}{lcccccccccccccccc}
    \toprule
    \multirow{4}[8]{*}{Method} & \multicolumn{2}{c}{\multirow{2}[4]{*}{Pre-training Data}} & \multicolumn{11}{c}{WER (\%)}                                   &     &     &  \\
\cmidrule{4-17}        & \multicolumn{2}{c}{} & \multicolumn{4}{c}{In-domain} & \multicolumn{5}{c}{Cross-domain} & \multicolumn{4}{c}{Cross-lingual} & \multirow{3}[6]{*}{AVG} \\
\cmidrule{2-16}        & \multicolumn{2}{c}{Librispeech} & \multicolumn{2}{c}{WSJ} & \multicolumn{2}{c}{AESRC} & \multicolumn{3}{c}{SWBD} & \multicolumn{2}{c}{TED} & \multicolumn{2}{c}{AISHELL-1} & \multicolumn{2}{c}{CV French} &  \\
\cmidrule{2-16}        & Labeled & Unlabeled & dev93 & eval92 & dev & test & RT03 & H-SB & H-CH & dev & test & dev & test & dev & test &  \\
    \midrule
    \textbf{1h fine-tune} &     &     &     &     &     &     &     &     &     &     &     &     &     &     &     &  \\
    \midrule
    Supervised Pre-train
        & 960h & ×   & 7.1 & 4.0 & 16.8 & 17.5 & 29.1 & 20.0 & 32.0 & 13.5 & 14.4 & 59.2 & 60.2 & 71.3 & 72.9 & 32.2 \\
    Wav2vec 2.0 & ×   & 960h & 8.4 & 6.4 & 16.0 & 16.8 & 28.1 & 19.9 & 28.9 & 17.1 & 15.1 & 67.3 & 66.8 & 61.0 & 63.4 & 31.9 \\
    Wav2vec-S & 100h & 860h & \textbf{5.4} & \textbf{3.8} & \textbf{11.3} & \textbf{10.9} & \textbf{22.6} & \textbf{14.2} & \textbf{22.7} & \textbf{10.0} & \textbf{9.9} & \textbf{48.9} & \textbf{48.7} & \textbf{51.2} & \textbf{53.9} & \textbf{24.1} \\
    \midrule
    \midrule
    \textbf{10h fine-tune} &     &     &     &     &     &     &     &     &     &     &     &     &     &     &     &  \\
    \midrule
    Supervised Pre-train
        & 960h & ×   & 6.2 & 3.6 & 13.5 & 13.6 & 25.8 & 15.6 & 29.7 & 12.2 & 12.8 & 27.0 & 27.8 & 46.8 & 49.9 & 21.9 \\
    Wav2vec 2.0 & ×   & 960h & 5.1 & 3.5 & 9.7 & 10.7 & 19.6 & 11.8 & 19.6 & 10.8 & 10.2 & 14.8 & 14.6 & 32.3 & 35.3 & 15.2 \\
    Wav2vec-S & 100h & 860h & \textbf{4.4} & \textbf{2.9} & \textbf{8.7} & \textbf{9.1} & \textbf{18.7} & \textbf{10.8} & \textbf{18.8} & \textbf{9.0} & \textbf{8.8} & \textbf{13.6} & \textbf{14.0} & \textbf{31.2} & \textbf{34.5} & \textbf{14.2} \\
    \bottomrule
    \end{tabular}%
    }
  \label{tab:main}%
\end{table*}%

\subsection{Main Results}
As shown in \autoref{tab:main}, we perform 1h and 10h fine-tuning on different pre-trained models. All pre-trained models are trained on the Librispeech but with different amounts of labeled/unlabeled data. The wav2vec 2.0 model trained on unlabeled data and the supervised pre-trained model trained on labeled data from scratch are used for comparison. 

Comparing the supervised pre-trained model and wav2vec 2.0 model, we find that for 1h fine-tuning, the supervised pre-trained model outperforms wav2vec 2.0 on 3 out of 6 datasets (WSJ, TED and AISHELL-1). However, when fine-tuning data increases to 10h, wav2vec 2.0 consistently outperforms the 960h supervised pre-trained model on all datasets. It illustrates the effectiveness and generalization of self-supervised pre-training. The proposed wav2vec-S model consistently outperforms the supervised and wav2vec 2.0 model on all datasets, demonstrating the effectiveness of the simple pipeline of wav2vec-S. Note that in \autoref{tab:main}, only 100h labeled data is used in wav2vec-S, although more labeled data could provide better results (shown in \autoref{sec:data}). Moreover, pre-training and fine-tuning both used character-level supervision and training updates is 20k. We will further discuss the impact of supervision level and training updates in \autoref{sec:supervision} and \autoref{sec:updates}. The following experiments are conducted with 10h fine-tuning on three representative datasets (WSJ, SWBD, AISHELL-1).

\subsection{Semi-supervised Pre-training data}
\label{sec:data}

\begin{table}[htbp]
  \centering
  \caption{Wav2vec-S performance with different semi-supervised pre-training data.}
  \resizebox{.48\textwidth}{!}{
    \begin{tabular}{cccccccccc}
    \toprule
    \multicolumn{2}{c}{\multirow{2}[4]{*}{Pre-training Data}} & \multicolumn{7}{c}{WER (\%)}            &  \\
\cmidrule{3-10}    \multicolumn{2}{c}{} & \multicolumn{2}{c}{WSJ} & \multicolumn{3}{c}{SWBD} & \multicolumn{2}{c}{AISHELL-1} & \multirow{2}[4]{*}{AVG} \\
\cmidrule(lr){1-2}   \cmidrule(lr){3-4}  \cmidrule(lr){5-7} \cmidrule(lr){8-9}     Labeled & Unlabeled & dev93 & eval92 & RT03 & H-SB & H-CH & dev & test &  \\
    \midrule
    100h & 0h  & 4.6 & 2.7 & 19.1 & 11.2 & 18.8 & 14.1 & 14.2 & 12.1 \\
    960h & 0h  & 4.3 & 2.6 & 19.0 & 10.8 & 18.6 & 13.5 & 13.8 & 11.8 \\
    100h & 860h & 4.4 & 2.9 & 18.7 & 10.8 & 18.8 & 13.6 & 14.0 & 11.9 \\
    \bottomrule
    \end{tabular}%
    }
  \label{tab:semi_data}%
\end{table}%

We compare using different amounts of labeled and unlabeled data during semi-supervised pre-training. As shown in \autoref{tab:semi_data}, the performance is the best when using all 960h labeled data and is the worst when using only 100h labeled data. Semi-supervised pre-training with 100h labeled and 860h unlabeled data effectively bridges the performance gap and achieves comparable performance with the 960h labeled one.

\subsection{Supervision Level}
\label{sec:supervision}

We discuss the optimal supervision level for semi-supervised pre-training. Specifically, We consider phone-level supervision and character-level supervision. The phoneme transcripts are generated using phonemizer\footnote{https://github.com/bootphon/phonemizer}.

\begin{table}[htbp]
  \centering
  \caption{Wav2vec-S performance with different supervision level for semi-supervised pre-training and fine-tuning.}
  \resizebox{.48\textwidth}{!}{
    \begin{tabular}{cccccccccc}
    \toprule
    \multirow{3}[5]{*}{Pre-train} & \multirow{3}[5]{*}{Fine-tune} & \multicolumn{7}{c}{WER (\%)}            &  \\
\cmidrule{3-10}        &     & \multicolumn{2}{c}{WSJ} & \multicolumn{3}{c}{SWBD } & \multicolumn{2}{c}{AISHELL-1} & \multirow{2}[3]{*}{AVG} \\
\cmidrule(lr){3-4}  \cmidrule(lr){5-7} \cmidrule(lr){8-9}        &     & dev93 & eval92 & RT03 & H-SB & H-CH & dev & test & \\ \midrule
    Phone & Phone & 5.9 & 4.7 & 20.2 & 13.1 & 20.2 & 15.8 & 15.3 & 13.6 \\
    Char & Phone & 5.6 & 4.8 & 19.9 & 13.2 & 20.2 & 15.9 & 15.3 & 13.6 \\
    Phone & Char & 4.8 & 3.3 & 19.3 & 11.3 & 19.1 & 14.7 & 15.5 & 12.6 \\
    Char & Char & 4.4 & 2.9 & 18.7 & 10.8 & 18.8 & 13.6 & 14.0 & 11.9 \\
    \bottomrule
    \end{tabular}%
    }
  \label{tab:supervision}%
\end{table}%

As shown in \autoref{tab:supervision}, when phone-level fine-tuning is used, phone-level and character-level pre-training perform similarly. On the other hand, when character-level fine-tuning is used, character-level pre-training clearly outperforms phone-level. It shows that the higher-level supervision (character) during pre-training can generalize well to the lower level (phone) but not vice versa. This conclusion also stands in the cross-lingual dataset (AISHELL-1), although the character supervision during pre-training and fine-tuning are in different languages. Therefore, we could conclude that character-level supervision is better for semi-supervised pre-training.

\subsection{Training Updates}
\label{sec:updates}

We illustrate the relation between the number of training updates and downstream performance in \autoref{tab:updates}.

\begin{table}[htbp]
  \centering
  \caption{Wav2vec-S performance with different training updates during semi-supervised pre-training. Valid denotes the validation WER on dev-other subset.}
  \resizebox{.45\textwidth}{!}{
    \begin{tabular}{cccccccccc}
    \toprule
    \multirow{3}[5]{*}{Updates} & \multicolumn{8}{c}{WER (\%)}                  &  \\
\cmidrule{2-10}        & \multirow{2}[3]{*}{Valid} & \multicolumn{2}{c}{WSJ} & \multicolumn{3}{c}{SWBD} & \multicolumn{2}{c}{AISHELL-1} & \multirow{2}[3]{*}{Avg} \\
\cmidrule(lr){3-4}  \cmidrule(lr){5-7} \cmidrule(lr){8-9}        &     & dev93 & eval92 & RT03 & H-SB & H-CH & dev & test &  \\ \midrule
    10k & 8.3 & 4.7 & 2.8 & 19.3 & 11.0 & 19.2 & 13.5 & 13.9 & 12.1 \\
    20k & 7.7 & 4.4 & 2.9 & 18.7 & 10.8 & 18.8 & 13.6 & 14.0 & 11.9 \\
    40k & 7.3 & 4.2 & 2.4 & 18.7 & 10.8 & 18.5 & 13.9 & 14.2 & 11.8 \\
    \bottomrule
    \end{tabular}%
    }
  \label{tab:updates}%
\end{table}%

When training updates increase, the WERs on source language datasets decrease, including in-domain (validation, WSJ) and the cross-domain (SWBD) datasets. On the contrary, the cross-lingual (AISHELL-1) WER increases. This indicates the trade-off between performances of the source language and other languages: with more training updates, the wav2vec-S model becomes more language-specific and the cross-lingual generalization ability is thus degraded.

\subsection{Training Time}

\begin{figure}[!htbp]
	\centering
	\includegraphics[width=0.8\columnwidth]{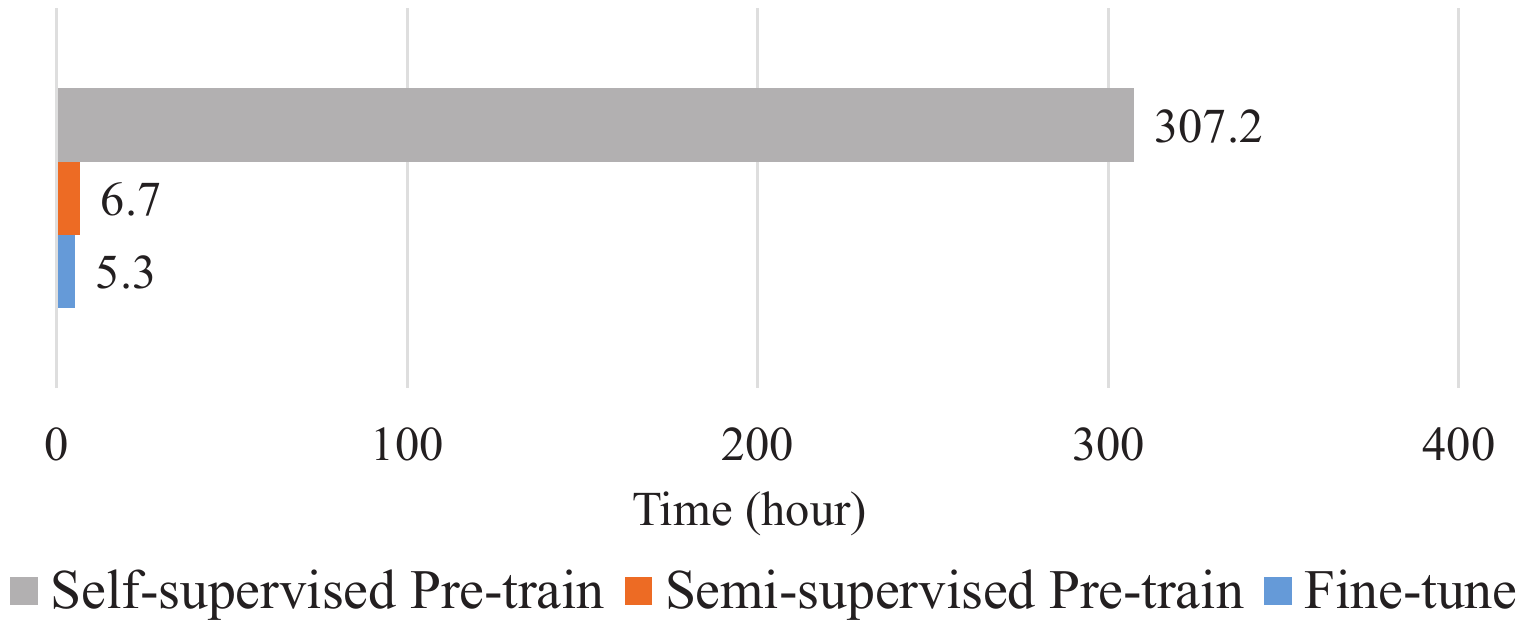}
	\caption{Comparison of training time for self-supervised, semi-supervised pre-training and fine-tuning.} 
	\label{fig:time}
\end{figure}

We conduct experiments using 8 V100 GPUs to show the training time for the two steps in wav2vec-S and fine-tuning in \autoref{fig:time}. The semi-supervised pre-training requires much less training time than the wav2vec 2.0 self-supervised training. The reason is that self-supervised pre-training can speed up the convergence of the followed semi-supervised pre-training. Moreover, since the self-supervised pre-training is task-agnostic, it can be reused by all downstream tasks. Therefore, for a new task, only semi-supervised pre-training is required to be conducted before fine-tuning.

\subsection{On Different Self-supervised Pre-trained Models}

\begin{table}[htbp]
  \centering
  \caption{Wav2vec-S performance with wav2vec 2.0 or data2vec as the self-supervised pre-training approach.}
  \resizebox{.45\textwidth}{!}{
    \begin{tabular}{lcccccccc}
    \toprule
    \multirow{3}[6]{*}{Method} & \multicolumn{7}{c}{WER (\%)}            &  \\
\cmidrule{2-9}        & \multicolumn{2}{c}{WSJ} & \multicolumn{3}{c}{SWBD} & \multicolumn{2}{c}{AISHELL-1} & \multirow{2}[4]{*}{AVG} \\
\cmidrule{2-8}        & dev93 & eval92 & RT03 & H-SB & H-CH & dev & test &  \\
    \midrule
    \textbf{1h fine-tune} &     &     &     &     &     &     &     &  \\
    \midrule
    wav2vec 2.0 & 8.4 & 6.4 & 28.1 & 19.9 & 28.9 & 67.3 & 66.8 & 32.3 \\
    wav2vec 2.0 + semi & 5.4 & 3.8 & 22.6 & 14.2 & \textbf{22.7} & 48.9 & 48.7 & 23.8 \\
    data2vec & 6.8 & 5.1 & 25.6 & 15.8 & 26.0 & 51.2 & 50.9 & 25.9 \\
    data2vec + semi & \textbf{5.3} & \textbf{3.4} & \textbf{22.6} & \textbf{13.3} & 22.8 & \textbf{45.9} & \textbf{45.5} & \textbf{22.7} \\
    \midrule
    \midrule
    \textbf{10h fine-tune} &     &     &     &     &     &     &     &  \\
    \midrule
    wav2vec 2.0 & 5.1 & 3.5 & 19.6 & 11.8 & 19.6 & 14.8 & 14.6 & 12.7 \\
    wav2vec 2.0 + semi & 4.4 & 2.9 & 18.7 & 10.8 & 18.8 & \textbf{13.6} & \textbf{14.0} & 11.9 \\
    data2vec & 4.6 & 3.0 & 19.2 & 10.7 & 19.2 & 14.0 & 14.2 & 12.1 \\
    data2vec + semi & \textbf{4.3} & \textbf{2.8} & \textbf{18.7} & \textbf{10.4} & \textbf{18.8} & 13.8 & 14.2 & \textbf{11.9} \\
    \bottomrule
    \end{tabular}%
    }
  \label{tab:data2vec}%
\end{table}%

We use another self-supervised pre-trained model data2vec to test the generalization of wav2vec-S on different self-supervised pre-trained models. As shown in \autoref{tab:data2vec}, data2vec outperforms wav2vec 2.0 on all datasets, illustrating the effectiveness of data2vec. And the wav2vec-S approach still consistently improves the performance of data2vec with the additional semi-supervised pre-training step. Therefore, wav2vec-S could act as a universal refinement approach to enhance a given self-supervised pre-trained model.

\subsection{Analysis with Representation Similarity}

We analyze the representation similarity before and after fine-tuning for different pre-trained models to show if a particular layer of the pre-trained model is suitable for the ASR task. Specifically, the fine-tuning is performed on the 10h subset. We follow the practice in \cite{pasad2021layer} and compute the CCA similarity between representations from each layer of a pre-trained model and the same layer of the corresponding fine-tuned model, where the lower CCA similarity means the representation changes more significantly during fine-tuning.

\begin{figure}[!htbp]
    \centering
	\subfloat[WSJ] 
	{ \label{fig:wsj}
		\includegraphics[width=0.75\columnwidth]{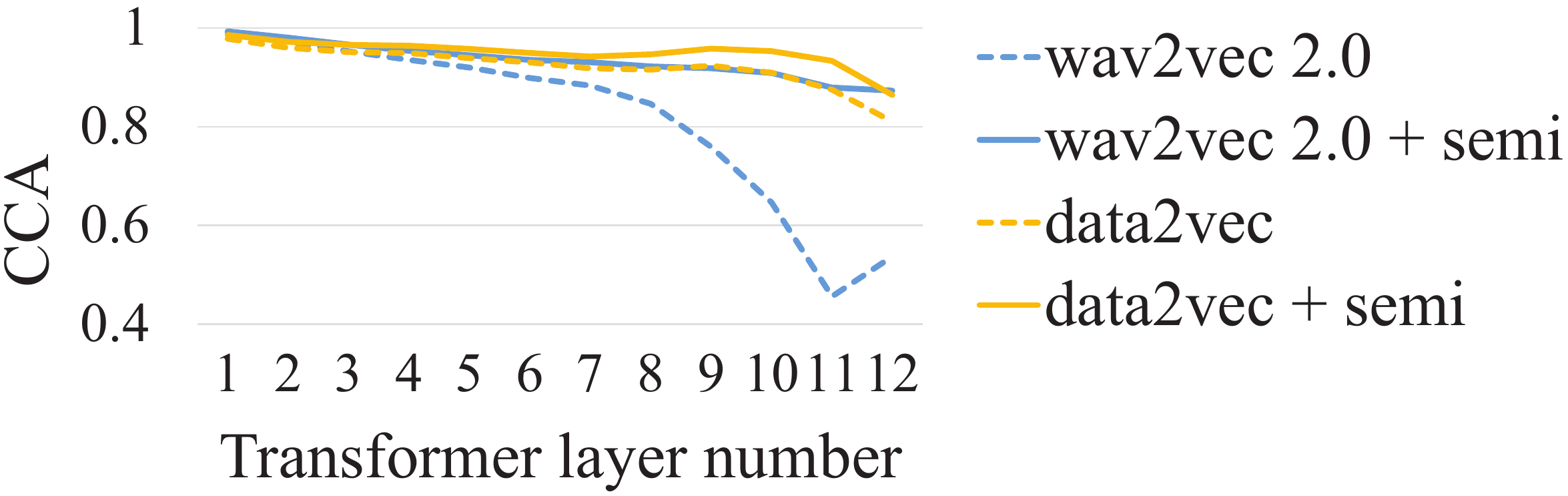}
	}		
	
	\subfloat[SWBD]  
	{ \label{fig:swbd}
		\includegraphics[width=0.45\columnwidth]{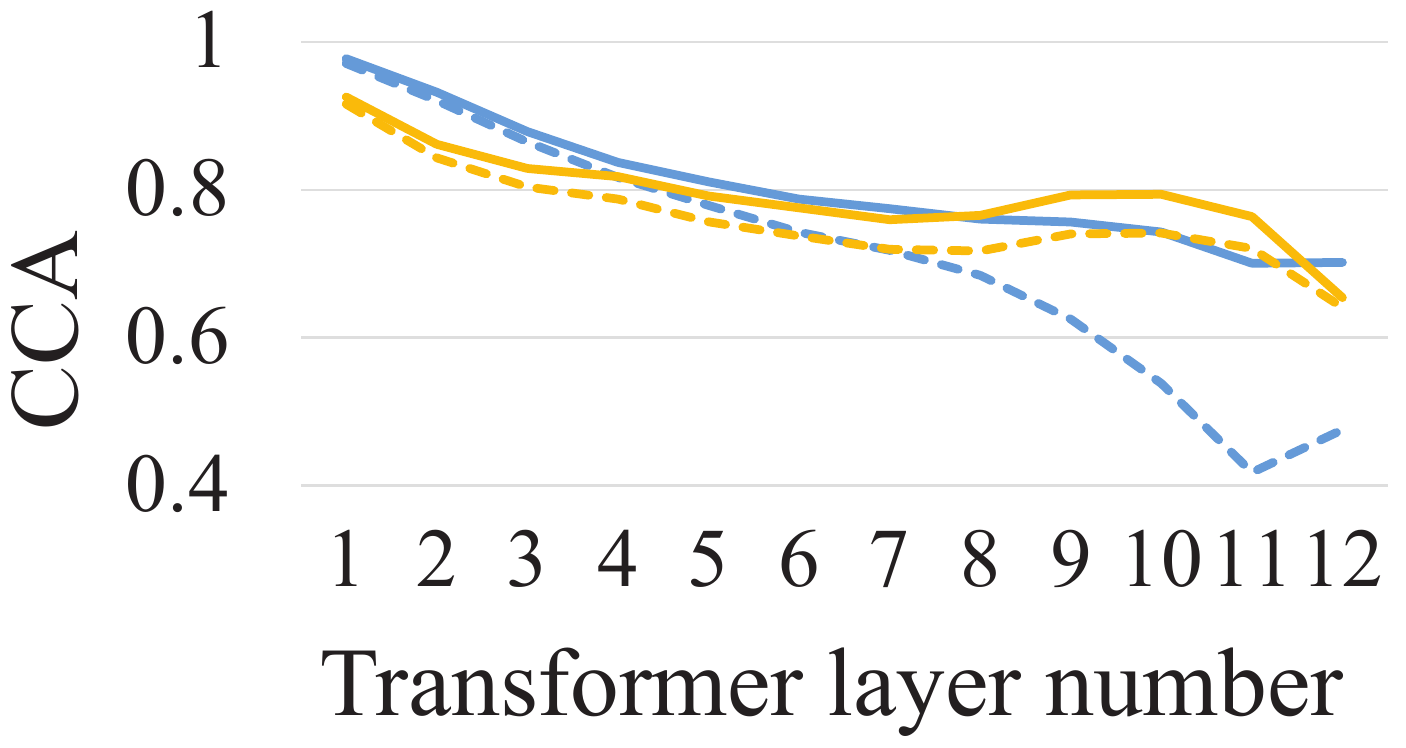}
	}		
	\thinspace
	\subfloat[AISHELL-1]  
	{ \label{fig:aishell}
		\includegraphics[width=0.45\columnwidth]{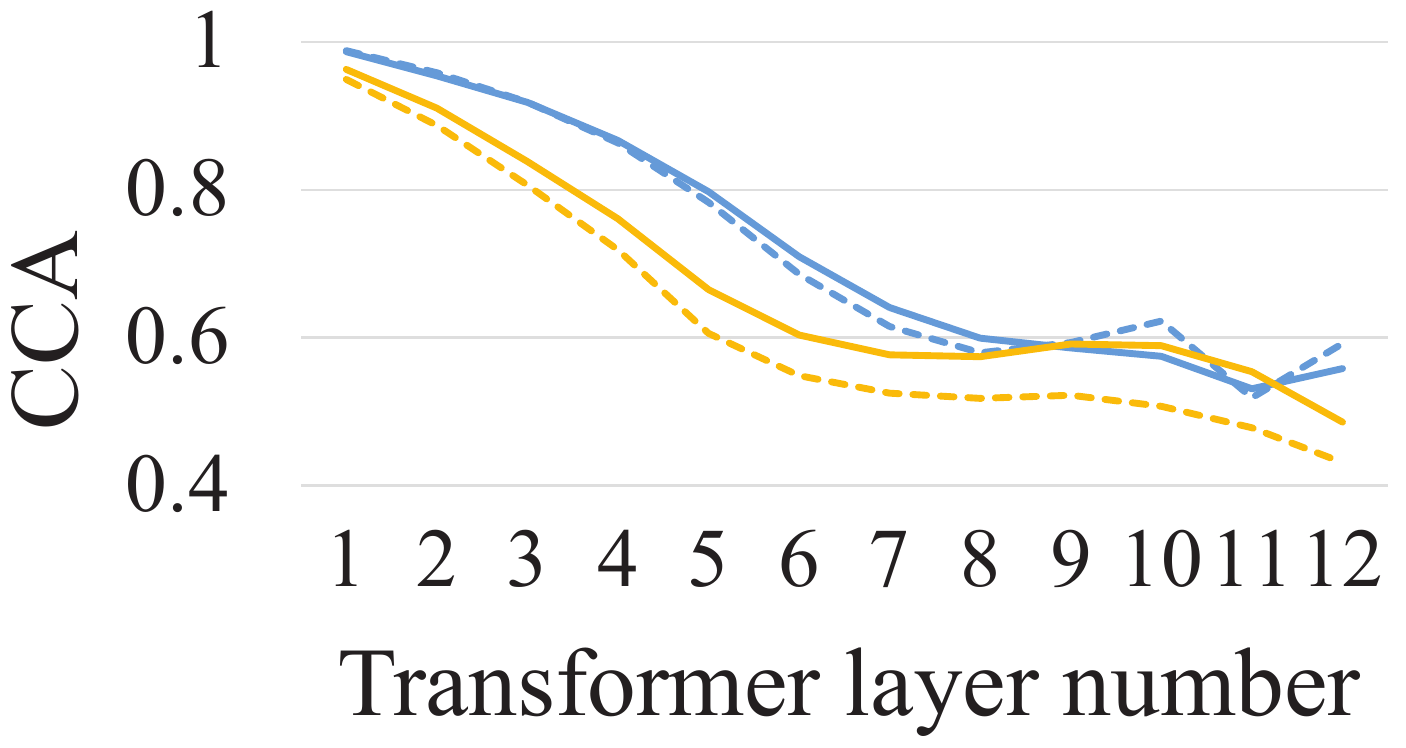}
	}		
	\caption{CCA similarity between each layer of a pre-trained
model and the same layer of corresponding fine-tuned model.} 
	\label{fig:cca}
\end{figure}

As shown in \autoref{fig:cca}, the last few layers of wav2vec 2.0 change significantly during fine-tuning, which illustrates the representations in its last few layers are less effective for ASR. This phenomenon is alleviated in data2vec on both in-domain and cross-domain datasets (WSJ and SWBD), which could be the reason why data2vec outperforms wav2vec 2.0. With the semi-supervised pre-training, both wav2vec 2.0 and data2vec have more similar representation to the fine-tuned model, illustrating semi-supervised pre-training effectively closes the representation gap between the task-agnostic self-supervised pre-trained model and task-specific fine-tuned model. 

There are some different phenomenons on the cross-lingual dataset AISHELL-1. Firstly, although data2vec performs better than wav2vec 2.0, it has lower CCA similarities in all layers. It means the CCA similarity could not directly reflect the ASR performance on the cross-lingual dataset, which is a significantly out-of-distribution dataset for pre-trained models. Secondly, the semi-supervised pre-training can not consistently improve the similarity in the last few layers of wav2vec 2.0. The reason might be that the labels in semi-supervised pre-training and fine-tuning are in different languages, which calls for the exploration of multi-lingual semi-supervised pre-training~\cite{lugosch2021pseudo}.

\section{Conclusions}
\label{sec:conclusion}

In this work, we propose wav2vec-S to build a better pre-trained model for low-resource ASR, which improves self-supervised pre-trained models via the task-specific refinement of semi-supervised pre-training. Experiments show that wav2vec-S consistently improves ASR performance on in-domain, cross-domain and cross-lingual datasets over self-supervised pre-trained models like wav2vec 2.0 and data2vec with a marginal increment of pre-training time.

\bibliographystyle{IEEEtran}
\bibliography{./main}

\end{document}